\documentclass[preprint,12pt]{aastex}

\usepackage{graphicx}
\usepackage{amsmath}
\usepackage{amssymb}
\usepackage{mathrsfs}

\begin{document}

\title{Testing the Predictions of the Universal Structured GRB Jet Model}

\author{Ehud Nakar\altaffilmark{1,2}, Jonathan Granot\altaffilmark{2} and Dafne Guetta\altaffilmark{1,3}}

\altaffiltext{1}{Racah Institute for Physics, The Hebrew
University, Jerusalem 91904, Israel} \altaffiltext{2}{Institute
for Advanced Study, Olden Lane, Princeton, NJ 08540}
\altaffiltext{3}{JILA, University of Colorado,
Boulder, CO 80309}


\begin{abstract}

The two leading models for the structure of GRB jets are the
uniform jet model and the universal structured jet (USJ) model. In
the latter, all GRB jets are intrinsically identical and the
energy per solid angle drops as the inverse square of the angle
from the jet axis. The simplicity of the USJ model gives it a
strong predictive power, including a specific prediction for the
observed GRB distribution as a function of both the redshift $z$
and the viewing angle $\theta$. We show that the current sample of
GRBs with known $z$ and estimated $\theta$ does not agree with the
predictions of the USJ model. This can be best seen for a
relatively narrow range in $z$, in which the USJ model predicts
that most GRBs should be near the upper end of the observed range
in $\theta$, while in the observed sample most GRBs are near the
lower end of that range. Since the current sample is very
inhomogeneous (i.e. involves many different detectors), it should
be taken with care and cannot be used to rule out the USJ model.
Nevertheless, this sample strongly disfavors the USJ model.
Comparing the prediction for the observed GRB distribution both in
$\theta$ and in $z$, with a larger and more homogeneous GRB
sample, like the one expected from Swift, would either clearly
rule out the USJ model, or alternatively, provide a strong support
for it. The test presented here is general, and can be used to
test any model that predicts both a luminosity function and a
luminosity-angle relation.

\end{abstract}

\section{Introduction}

Since the discovery of gamma-ray burst (GRB) afterglows in early
1997, several lines of evidence have emerged in support of
collimated outflows, or jets. 
The total energy output in $\gamma$-rays that is
inferred from the fluence of bursts with a known redshift $z$,
assuming spherical symmetry, $E_{\rm \gamma,iso}$, in some cases
approaches and in one case (GRB 990123) even exceeds $M_\odot c^2$.
This is hard to produce in any progenitor model involving a
stellar-mass object. A jet can significantly reduce the total energy
output for a given $E_{\rm \gamma,iso}$.  A more direct line of
evidence in favor of jets in GRBs (and probably the best
evidence so far), is from achromatic breaks in the afterglow light
curves (Rhoads 1997,1999; Sari, Piran \& Halpern 1999).

However, despite six years of
extensive afterglow observations, the structure of the
relativistic jets that produce GRBs is still unknown. The
structure of GRB jets is a very fundamental and important property
which effects the requirements from the source that accelerates
and collimates the jets, and has direct bearing on two of the most
basic properties of any astrophysical radiation source: the rate
and the total amount of energy output. Within the fireball model
(for review see Piran 2000, M\'esz\'aros 2002) there are two very
different jet structures which are compatible with the
observations: (i) the uniform (or `top hat') jet model (Rhoads
1997,1999; Panaitescu \& M\'esz\'aros 1999; Sari, Piran \& Halpern
1999; Kumar \& Panaitescu 2000; Moderski, Sikora \& Bulik 2000;
Granot et al. 2001,2002), where the initial energy per solid angle
$\epsilon$ and Lorentz factor $\Gamma$ are uniform within some
finite half-opening angle $\theta_j$ and sharply drop outside of
$\theta_j$, and (ii) the universal structured jet (USJ) model
(Postnov, Prokhorov \& Lipunov 2001; Rossi, Lazzati \& Rees 2002;
Zhang \& M\'esz\'aros 2002), with a standard jet structure for all
GRBs where $\epsilon \propto \theta^{-2}$ (outside of some core
angle). It is important to note that in the USJ model all GRB jets
are intrinsically identical (both in their angular profile and
total energy). Both jet structures can explain the observed
correlation between $E_{\rm\gamma,iso}$ and the jet break time $t_j$
in the optical afterglow light curve, $t_j\propto
E_{\rm\gamma,iso}^{-1}$ (Frail et. al. 2001; Bloom et al. 2003).
In the USJ model this  determines the jet structure ($\epsilon
\propto \theta^{-2}$). In the `top hat' model $t_j$ depends mainly
on $\theta_j$ (Rhoads 1997,1999; Sari et. al. 1999), while in the
USJ model it depends mainly (and in a similar way) on the viewing
angle $\theta_{\rm obs}$ from the jet axis.

The simplicity  the USJ model gives it a strong predictive power.
In a recent paper Perna, Sari \& Frail (2003; hereafter PSF03)
used this feature of the USJ model to predict the observed
distribution of viewing angles, $n(\theta)=dn/d\theta$ (hereafter
we use $\theta$ instead of $\theta_{\rm obs}$ for brevity). They
have shown that the current limited sample of $16$ bursts with
known $\theta$ and redshift $z$ fits the predicted distribution
very well. In this Letter we extend the comparison between the USJ
model predictions and observations into one more dimension - the
redshift $z$. Namely, we use here the two dimensional (2D)
distribution $n(z,\theta)=dn/dzd\theta$ and compare it with the
known $\theta$ and $z$ of the observed sample. The 1D distribution
that was used by PSF03 for comparison with the data is $n(\theta)
=\int n(z,\theta) dz$. However, the known redshift of these GRBs
is not used in the 1D analysis. Comparing the data both for $z$
and for $\theta$ with the 2D distribution, $n(z,\theta)$, reveals
that the agreement between the data and the 1D distribution,
$n(\theta)$, that was found by PSF03 is accidental, and arises
because of the integration over the z. The 2D data shows a very
poor agreement with the model, and the hypothesis that the data is
drawn from the model is rejected at $99\%$ significance by a 2D
K-S test. Thus, the agreement between the data and the 1D
distribution is misleading, and does not provide support for (and
certainly does not prove) the USJ model. In order to test this
statistically, while trying to minimize the possible selection
effects as a function of $z$, it is most appropriate to compare
$n(z,\theta)$ with the data over a relatively narrow range in $z$.
Such a comparison has another important advantage: it depends only
weakly on $R_{\rm GRB}(z)$ - the GRBs rate as a function of $z$ -
which is rather poorly known. A drawback of this test is that it
requires a large number of data points at a given redshift. The
narrowest range in $z$ that contains a reasonably large number of
the current data points (10 points) is $0.8<z<1.7$. In this range
the USJ model fails to explain the data with a significance of
$99.8\%$.

We perform several tests in order to check the robustness of this
result, and its sensitivity to various selection effects that we
can quantify, and find it to be robust and significant. Thus, we
find that the current data set strongly disfavors the USJ model.
Nevertheless, it is still premature to draw a definite conclusion,
mainly because  many different detectors were involved in
detecting the current sample. This situation is expected to
improve in the near future with the launch of Swift. Once a
homogeneous and large sample of bursts with measured $z$ and
estimated $\theta$ would be available, applying the 2D test
described here would result in either a definite rejection or a
strong support for the USJ model. We stress that the specific test
we carry here is relevant only to a {\it universal} structured
jet, i.e. with a universal profile of both $\epsilon(\theta)$ and
the initial value of $\Gamma(\theta)$.\footnote{if the initial
$\Gamma$ profile is  not universal, then different jets may
produce $\gamma$-rays within different solid angles.} Finally,
although in this Letter we apply the 2D test only to the USJ
model, it can be easily generalized to any model that predicts a
luminosity function and a luminosity-angle relation.

\section{Theory}
\label{theory}

Below we follow PSF03, and generalize their 1D distribution
$n(\theta)$ to the 2D distribution $n(z,\theta)$. Eq. 5 of PSF03
presents the photon peak luminosity in the energy range
$50-300\;$keV, assuming that all GRBs are identical with a
differential photon spectral index, $\alpha=1$,
\begin{equation}
 \label{EQ Lph} L_{\rm ph}(\theta,T)=1.1 \times 10^{57}\tilde{T}^{-1}
 \theta^{-2}\;{\rm photons\;sec^{-1}}\ ,
\end{equation}
where $T=\tilde{T}\;$sec is an ``effective" duration which is
given by the ratio of the (isotropic equivalent) energy output and
peak luminosity (or equivalently $(1+z)^{-1}$ times the ratio of
the fluence and peak flux). In practice, $T$ changes from one
burst to another, and we will denote its probability distribution
by $P(T)$. For a detector with a given limiting flux for detection
$F_{\rm ph,lim}=\tilde{F}_{\rm
  ph,lim}\;{\rm photons\;cm^{-2}\;sec^{-1}}$ and a burst at given
$\theta$ and $z$, we can derive the maximal $T$ for which this burst
is detected, $T_{\rm max}$,
\begin{equation}\label{EQ T_max}
T_{\rm max}=88\,(1+z)^{-\alpha}D_{28}^{-2}(z)\tilde{F}_{\rm
ph,lim}^{-1}\left(\frac{\theta}{0.1}\right)^{-2}\;{\rm sec}\ ,
\end{equation}
where $D_{28}(z)$ is the co-moving distance in units of
$10^{28}\;$cm. The total rate of bursts with an inferred viewing
angle between $\theta$ and $\theta+d\theta$ and a redshift between
$z$ and $z+dz$ is then
\begin{equation}
 \label{EQ n(theta_z)} n(z,\theta)=\frac{dn}{dzd\theta} =
 \sin\theta\frac{R_{\rm GRB}(z)}{(1+z)} \frac{dV(z)}{dz}
\int_0^{T_{\rm max}(\theta,z)}P(T)dT \ ,
\end{equation}
where $R_{\rm GRB}(z)$ is the GRB rate per unit comoving time per
unit comoving volume $V(z)$. Eq.  \ref{EQ  n(theta_z)} is similar
to Eq. 11 of PSF03 but without integration over $z$. Thus, Eq.
\ref{EQ n(theta_z)} describes the 2D distribution $n(z,\theta)$
while Eq. 11 of PSF03 describes the 1D distribution
$n(\theta)=dn/d\theta$.

In order to calculate $n(z,\theta)$ one must assume some $R_{\rm
  GRB}(z)$ and $P(T)$. Below we consider the $R_{\rm GRB}(z)$ which
PSF03 used as their ``standard" model\footnote{This model is the
  Rowan-Robinson (1999) star formation rate with a cutoff at $z>10$ as
  seen in the numerical simulation of Gnedin \& Ostriker (1997)} where
for $z<10$:
\begin{equation}
 \label{EQ R_GRB} R_{\rm GRB}(z) \propto \left\{\begin{array}{c}
   10^{0.75z}, \ \ \ \ \ \ \ z<z_{\rm peak}\\
   10^{0.75z_{\rm peak}}, \ \ \ \ z\geq z_{\rm peak}
 \end{array} \right.,
\end{equation}
and $z_{\rm peak}=2$. At $z>10$ the rate declines rapidly. PSF03
first considered a delta function in $T$, $P(T)=\delta(T-T_0)$,
where $T_0$ is a free parameter which they used in order to get a
good fit to the observed $dn/d\theta$. Using their ``standard"
$R_{\rm GRB}(z)$, they found a best fit value of $T_0=8\;$sec.
However, since $P(T)$ can be estimated pretty well from
observations, we do not take it as a free function. In order to
estimate $P(T)$ we used the flux table of the BATSE 4B-catalog. We
have used the peak fluxes (in ${\rm
  photons\;cm^{-2}\;sec^{-1}}$), averaged over the $1024\;$ms BATSE
trigger, and the fluences from the catalog in the energy range
$50-300\;$keV.  In order to convert the peak fluxes to ${\rm
  erg\;cm^{-2}\;sec^{-1}}$ we used a Band spectrum (Band et al.  1993)
for the energy distribution, with $\alpha=-1.0$, $\beta=-2.0$ and
$E_0=100\;$keV. Finally, $T$ is approximated by the ratio of the
fluence and the peak flux.\footnote{This ratio includes the effect of
  cosmological time dilation, and is therefore a factor of $(1+z)$
  larger than the actual value of $T$, which is measured at the
  cosmological frame of the GRB. Therefore, we overestimate both
  $\langle T\rangle$ (by a factor of $\sim 2-3$) and $\sigma_{\ln T}$
  (as part of the observed scatter is due to the scatter in $(1+z)$
  between different GRBs). Both affects worsen the fit between the USJ
  model and the data.
  }
According to this estimation of $T$, 
the total distribution $P(T)$ is consistent with a lognormal
distribution,
\begin{equation}\label{EQ_P(T)}
\frac{dP}{d\ln\tilde{T}}=\tilde{T}P(\tilde{T})=\frac{1}{\sigma_{\ln T}
\sqrt{2\pi}} \exp\left[-\frac{(\ln\tilde{T}-\mu)^2}{2\sigma_{\ln T}^2}\right]\
,
\end{equation}
with $\mu = 2.15$ and $\sigma_{\ln T}=0.87$ ($T=8.6 \pm
^{12}_{5}$sec). Below we first use the delta function for $P(T)$
that was used by PSF03, and later we use our Eq. \ref{EQ_P(T)}.

Next we consider measurement errors or intrinsic scatter in
$\epsilon\,\theta^2$. The measurement errors in $z$ are typically
negligible. The error in $\theta$, however, may be as large as
tens of percent, and so is the intrinsic scatter in
$\epsilon\,\theta^2$. In order to account for this scatter, and
since it is more reasonable to assume a Gaussian scatter in
$\ln\theta$ rather than in $\theta$ (both for error and intrinsic
scatter), we change variables from $\theta$ to $\ln\theta$, and
convolve $n(z,\ln\theta)=dn/dzd\ln\theta=\theta
dn/dzd\theta=\theta\, n(z,\theta)$ along the $\ln\theta$
coordinate with a Gaussian of standard deviation
$\sigma_{\ln\theta}$,
\begin{equation}
 \label{EQ n_tilde} \tilde{n}(z,\ln\theta)=
 \frac{d\tilde{n}}{dz d\ln\theta}=
 \frac{\int_0^{\ln(\pi/2)}n(z,\ln\theta')\exp\left[
 -(\ln\theta-\ln\theta')^2/2\sigma_{\ln\theta}^2\right] d(\ln\theta')}
 {\int_0^{\ln(\pi/2)}
 \exp\left[-(\ln\theta-\ln\theta')^2/2\sigma_{\ln\theta}^2\right]
 d(\ln\theta')}\ .
\end{equation}
Now $\tilde{n}(z,\ln\theta)$ is a rate function which is smoothed
along the $\ln\theta$ dimension by the typical scatter due to the
bursts intrinsic properties and $\theta$ measurement error. The total
scatter cannot exceed the measured scatter in $\epsilon\,\theta^2$
that was found by Frail et al. (2001).

\section {Results}
\label{results}

First we repeat the 1D analysis of PSF03 in 2D for their ``standard"
$R_{\rm GRB}(z)$ model (Eq. \ref{EQ R_GRB}), using
$P(T)=\delta(T-T_0)$ with $T_0=8\;$sec, for which they obtained the
best fit to the data. We used the same parameters as PSF03: $F_{\rm
  ph,lim}=0.424\;{\rm photons\;cm^{-2}\;sec^{-1}}$ (The threshold for
the BATSE trigger on $1024\;$ms; Mallozzi, Pendleton \& Paciesas
1996), $\alpha=1$, and the same cosmology: $\Omega_M=0.3$,
$\Omega_\Lambda =0.7$, and $H_0 = 71\;{\rm
km\;sec^{-1}\;Mpc^{-1}}$.

\begin{figure}[h]
\begin{center}
\includegraphics[width=17cm,height=10cm]{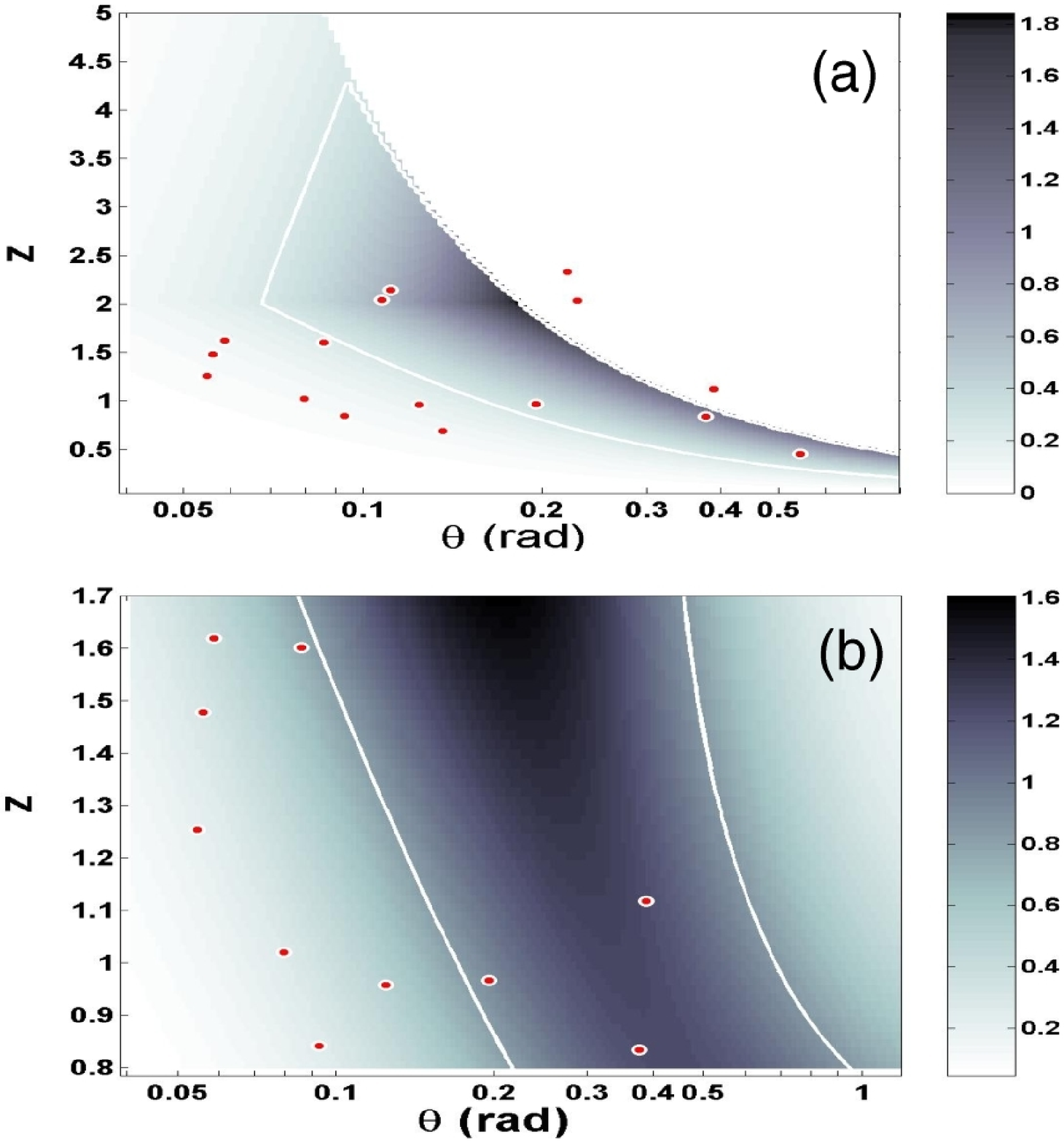}
\caption{The 2D distribution density, $n(z,\ln\theta)$, of the GRB
  rate as a function of redshift $z$ and viewing angle $\theta$, as
  predicted by the universal structured jet (USJ) model.
  The white contour lines confine the minimal
  area which contains $1\;\sigma$ of the total probability. The circles
  denote the $16$ bursts with  known $z$ and $\theta$ from the sample of
  Bloom et al. (2003). {\bf (a)}: The parameters of the models are similar
  to these of PSF03. This figure is the 2D realization of their Fig.
  1. {\bf (b)} Here we use a limited range in redshift, $0.8<z<1.7$
  (containing $10$ out of the $16$ data points), in order to minimize
  redshift selections effects and reduce the sensitivity of the results
  to the unknown GRB rate. We take into account $20\%$ measurement errors
  in $\ln\theta$ ($\sigma_{\ln\theta}=0.2$) and a lognormal distribution
  in $T$ that we deduced from observations (Eq. \ref{EQ_P(T)}).}
 \label{plotone}
\end{center}
\end{figure}

Fig. \ref{plotone}a depicts the 2D distribution $n(z,\ln\theta)$,
and the circles mark the $16$ GRBs of the Bloom et al. (2003)
sample, which were used by PSF03. This figure is a 2D
representation of Fig. 1 of PSF03. When integrating our Fig.
\ref{plotone}a over the $z$ dimension we reproduce Fig 1 of PSF03
(model 1). Fig. \ref{plotone}a shows that while the 1D
distribution ,$n(\theta)$, provides a good fit to the data, the 2D
distribution, $n(z,\theta)$, does not agree with the data. The 2D
K-S test rejects the null hypothesis that the data is drawn from
the model with a confidence of $99\%$.  The reason for the
striking difference between the 2D analysis and the 1D analysis is
that the data disagree with the 2D model in two ways that roughly
cancel out when integrated over redshift: (i) at high $z$
($\gtrsim 2$) there are not enough bursts with low $\theta$ , and
(ii) at low $z$ there are too many burst with low $\theta$
compared to the number of bursts with high $\theta$ ($\gtrsim
0.2\;$rad). When integrating over the redshift these two
shortcomings roughly cancel each other out. The fact that the 1D
distribution of PSF03 peaks at $\theta\sim 0.12\;$rad arises from
contribution of predicted bursts at high $z$ and low $\theta$,
which are not present in the observational sample, but are
compensated for by the overabundance of bursts at low $z$ and low
$\theta$.

This disagreement with the data, however, cannot be used to draw
strong conclusions. The reason is that the current sample suffers
from numerous selection effects, mainly in redshift. 
The selection effects in $z$ can be minimized by testing the
$\theta$ distribution for a given $z$ (i.e $dn/d\theta$ for a
given $z$). This test has another important advantage: it depends
only weakly on the poorly known GRB rate $R_{\rm
  GRB}(z)$. The main disadvantage of this test is that the size of the
data sample is reduced. We therefore take a slice in redshift of
$0.8<z<1.7$, which contains $10$ of the $16$ bursts in the current
sample. We use $R_{\rm GRB}(z)$ and $P(T)$ from Eqs. \ref{EQ
R_GRB} and \ref{EQ_P(T)}. We account for a $20\%$ scatter in
$\ln\theta$ by using $\tilde{n}(z,\ln\theta)$ from Eq. \ref{EQ
n_tilde} with $\sigma_{\ln\theta}=0.2$. Fig. \ref{plotone}b shows
the expected distribution $\tilde{n}(z,\ln\theta)$ in this
redshift range. Here the paucity of bursts with large $\theta$,
and overabundance of bursts with small $\theta$, is clear. The
concentration of bursts at $\theta<0.1$ while $\theta_{\rm max}(z)
\sim 0.25-0.4$ in this $z$ range contradicts the predictions of
the USJ model - $n(\theta) \propto \sin\theta$, for
$\theta<\theta_{\rm max}$. The 2D K-S test rejects the model with
a confidence level of $99.8\%$.

In order to check the reliability and robustness of our results,
we consider below the sensitivity of the results to our
assumptions and to the values of the different parameters ($z_{\rm
peak}$, $F_{\rm ph,lim}$, etc.). We also consider different
observational selection effects and carry additional tests to
estimate their influence on the results. First, we varied the
value of $z_{\rm peak}$ (Eq. \ref{EQ R_GRB}), and found that it
has almost no influence on the results. We considered also the
possibility of a larger $\sigma_{\ln\theta}$.\footnote{As
discussed above it accounts for both the intrinsic scatter and
measurement errors.} Bloom et al. (2003) obtain a factor of $2.2$
scatter in $\epsilon\,\theta^2$, implying
$\sigma_{\ln\theta}\lesssim 0.4$.  Repeating our analysis with
$\sigma_{\ln\theta}=0.4$, the USJ model is rejected at $98.7\%$
confidence.

Next we consider the dependence of our result on the value of
$F_{\rm ph,lim}$. It is important to note that we consider here a
stiff threshold (i.e. constant $F_{\rm ph,lim}$). In reality it is
not the case (both the detection threshold of the detector and the
level of the background vary). This effect may result in an
underestimate of the number of weak events. One way to overcome
this obstacle is by choosing a relatively bright threshold which
is above the detection threshold at any time.  Unfortunately, this
significantly reduces the size of the observed sample, preventing
us from applying this method here (it may be applied to a future
larger sample, e.g. Swift). The only test we can do with the
current sample is to check the effect of a larger (stiff)
threshold. Lower sensitivity (larger $F_{\rm ph,lim}$) reduces
$\theta_{\rm  max}(z)$. In order for the model to accommodate the
data concentration at $\theta<0.1\;$rad and $0.8<z<1.7$, $F_{\rm
ph,lim}$ needs to be increased by a factor of $\sim 5$. However,
this would imply $\theta_{\rm max}(z=2)\approx 0.06\;$rad, which
is significantly inconsistent with the observational values of
$\theta$ at this redshift (two bursts with $\theta \approx 0.12$
and two with $\theta \approx 0.22$). Another threshold related
selection effect is the low sensitivity of BATSE in the X-ray.
This effect is important because of the correlation $E_p \propto
\epsilon^{1/2}$ (Amati et. al 2002), which in our context implies
$E_p \propto \theta^{-1}$. Thus, BATSE is less sensitive to bursts
with large $\theta$ (low $E_p$). Five bursts from our specific
sample are also used by Amati et al. (2002): four with $\theta <
0.1\;$rad and $E_p \gtrsim 400\;$keV, and one (GRB 970508) with
$\theta=0.38\;$rad and an intrinsic $E_p \sim 150\;$keV. This
sub-sample roughly follows the relation $E_p \propto \theta^{-1}$,
and demonstrates that in the range of $\theta$ where there is a
deficit of observed bursts, $\sim 0.2-0.3\;$rad, the expected
intrinsic $E_p$ is $\sim 200\;$keV (observed $E_p\sim 100\;$keV),
which is well within the range of BATSE. Therefore, although we
cannot quantify this effect accurately, it should not strongly
affect our sample with limited $z$ range.

Next we consider the selection effects in $\theta$. A very small
$\theta$ implies a very early jet break time $t_j$, which may be
before the first optical detection and thus result only in an
upper limit on $\theta$. A large $\theta$, on the other hand,
results in a late jet break time $t_j$, which occurs when the
optical afterglow is too dim for detection (it can be either
dimmer than the detection threshold or its host galaxy). This
would result in only a lower limit on $\theta$. The sample of
Bloom et al. (2003) shows both effects: in the redshift range
$0.8<z<1.7$ there is one burst with an upper limit on $\theta$ and
two with lower limits on $\theta$. In order to account for the
above selection effects, we added the latter three bursts in the
most favorable way for the USJ model. Namely, we assigned to each
of the three bursts the value of $\theta$ within the allowed range
where $\tilde{n}(\theta,z)$ assumes its maximal value. Even after
adding these three data points in this way, the 10+3 data set
rejects the USJ model with a confidence level of $99.8\%$.

Another selection effect in $\theta$ may result from ``dark"
bursts. These are bursts with an observed X-ray afterglow in which
an optical afterglow was not detected despite deep and rapid
follow-up observations. De Pasquale et al. (2003) argue that a
large fraction of the ``dark" bursts are intrinsically dim, by
showing that they have, on average, dimmer X-ray afterglows. In
the USJ model the intrinsically dim bursts may be interpreted as
bursts with very large $\theta$ which are barley detected in
$\gamma$-rays and X-rays. Since about $50\%$ of the bursts with
observed X-ray afterglows have no detected optical afterglow, we
will make the extreme assumption that all of these bursts are
intrinsically dim (i.e. have a large $\theta$). We estimate the
effect of this assumption by adding 13 fictitious points (to the
$10+3$ original data points) in the most favorable way for the
model. Namely, we generated $z$ according to the model $z$
distribution, and then for each point we choose $\theta$ where
$\tilde{n}(\theta,z)$ is maximal (at the given $z$ of that point).
Even after making these extreme assumptions, the 26 ``data" points
reject the model with confidence level larger than $90\%$. We
conclude that our result, that the USJ model is incompatible with
the current data, is robust and significant.

\section{Discussion}
\label{diss}

We have shown that the strong predictive power of the universal
structured jet (USJ) model enables a determination of the expected
distribution of observed GRB rate as a function of both redshift
$z$ and viewing angle $\theta$. We have compared this predicted 2D
distribution to current observations, and found a very poor
agreement. This is in contrast with the result of PSF03 which
compared only the observed distribution of $\theta$ to the 1D
prediction of the USJ model (that is obtained from the 2D
prediction by integrating over $z$) and found a good agreement.
Our analysis shows that this agreement is accidental (resulting
from the integration over $z$) and does not support the USJ model
in any way.

However, the poor agreement between the data and the 2D
distribution should be taken with care, and may not be used to
draw definite conclusions. This is since the current GRB sample is
highly non-homogeneous. It involves many different instruments,
and is likely effected by various selection effects, which are
hard to quantify very accurately. A larger and much more
homogeneous sample of GRBs with known $z$ and $\theta$ is expected
with the upcoming launch of Swift, which would enable much
stronger and clearer conclusions to be drawn from a similar
statistical analysis as was done in this work.\footnote{Note that
the effect of the variable detection threshold and the  $E_p -
\epsilon$ correlation, should be considered carefully in the
analysis of Swift results as well.} This would either clearly rule
out or strongly support the USJ model. Nevertheless, we point out
that at least some of the selection effects may be overcome by
restricting the analysis to a relatively narrow range in $z$. This
significantly reduces the redshift selection effects and the
uncertainty that is introduced by the assumption that has to be
made about the poorly known GRB rate $R_{\rm
  GRB}(z)$. The main drawback of this second method is that in order
to obtain a statistically significant sample in a very narrow
range in $z$, many GRBs (many more than in the current sample) are
required. Again, such a sample is expected to become available
with Swift.

In the current sample, the narrowest redshift range which still
contains enough data points for the purpose of statistical
analysis is $0.8<z<1.7$ (it contains $10$ out of the $16$ points).
In this range, the current data is in complete disagreement with
the predictions of the USJ model, with a significance of $99.8\%$
according to the 2D K-S test. We checked the robustness of this
result by varying our assumptions and by examining a few possible
selection effects in the viewing angle $\theta$, and found it to
be robust. Thus, although the relatively small size and the
inhomogeneity of the current sample prevent us from drawing a
definite conclusion at this stage, we find that the current data
disfavor the USJ model.

\acknowledgements

We are grateful to Rosalba Perna for her comments about the selection
effects related to the viewing angle, which greatly improved the
paper. We thank Tsvi Piran for illuminating discussions and helpful
remarks. EN is supported by the Horowitz foundation and through the
generosity of the Dan David Prize Scholarship 2003. EN thanks the
Institute for Advanced Study for its warm hospitality and great
working atmosphere during the course of this work. JG is supported by
the W.M. Keck foundation, and by NSF grant PHY-0070928. DG
acknowledges the RTN ``Gamma-Ray Bursts: An Enigma and a Tool''and NSF
grant AST-0307502 for supporting this work.

\clearpage

\end{document}